\documentclass[11pt]{article}
\usepackage{moriond,epsfig,subfigure}

\def\Journal#1#2#3#4{{#1} {\bf #2}, #3 (#4)}

\def\EPJ{{\em Eur. Phys. J.} C}
\newcommand{\pom} {I\hspace{-0.2em}P}
\newcommand{\xpom}{\mbox{$x_{_{\pom}}$}}
\newcommand{\zpom}{\mbox{$z_{_{\pom}}$}}
\newcommand{\zpomjet}{\mbox{$z_{_{\pom}}^{\rm jets}$}}

\def\be{\begin{equation}}
\def\ee{\end{equation}}
\def\bea{\begin{eqnarray}}
\def\eea{\end{eqnarray}}

\begin{document}
\vspace*{4cm}
\title{Diffraction at HERA}

\author{ Heuijin Lim (on behalf of H1 and ZEUS collaborations)} 

\address{ High Energy Physics Division, Argonne National Laboratory, \\
9700 S. Cass Avenue, Argonne, IL 60439, USA}

\maketitle\abstracts{
Precision measurements of diffraction have been performed by the H1 and 
ZEUS experiments at the HERA collider with high statistics for a wide  
kinematic range of photon virtuality $Q^2$. The diffractive parton densities
are extracted by performing the NLO DGLAP QCD fits to diffractive data
and can be used to test QCD factorisation with diffractive final states.
%
}

\section{Introduction}
\vspace*{-0.1cm}
Diffraction, in which the proton or
a low-mass nucleonic system emerges from the interaction with almost the full
energy of the incident proton, is mediated by the exchange of a colour singlet
carrying the quantum numbers of the vacuum, called the Pomeron. It has been
observed by the 
presence of a large rapidity gap between the proton and the rest of the final 
state, which is not exponentially suppressed.

With increasing statistics, improved instruments and better detector
understanding, the H1 and ZEUS experiments at HERA have measured the 
diffractive processes for many different final states. 
Diffractive measurements at HERA are necessary for understanding the low-$x$
structure of the proton and also important for the interpretation of LHC
results.


\section{Measurements of inclusive diffraction}
\vspace*{-0.1cm}
Inclusive diffractive events have been extracted using the hadronic mass
spectrum
observed in the central detector ($M_X$ method~\cite{FPCII}), the presence of
a large rapidity gap (LRG method~\cite{H1_Etamax,ZEUS_EtaLPS_Pre}), or the detection of the leading protons which carry a large fraction
of the incoming proton beam energy (H1 FPS~\cite{H1_LPS}, ZEUS LPS methods~\cite{ZEUS_EtaLPS_Pre}).
While additional contributions from Reggeon
exchange are exponentially suppressed when using the $M_X$ method,
the selections based on LRG or on a
leading proton may include these contributions.
The LPS method can reconstruct the squared four-momentum transfer at the
proton vertex $t$ under the limited LPS acceptance. The other two methods
based on
the hadronic activity in the forward detector only provide the results
integrated over $t$ and therefore contain the contribution from the
nucleon dissociation to $M_N < 2.3$ GeV ($M_N < 1.6$ GeV) for ZEUS (H1)
measurement.

\begin{figure}
  \begin{center}
    \psfig{figure=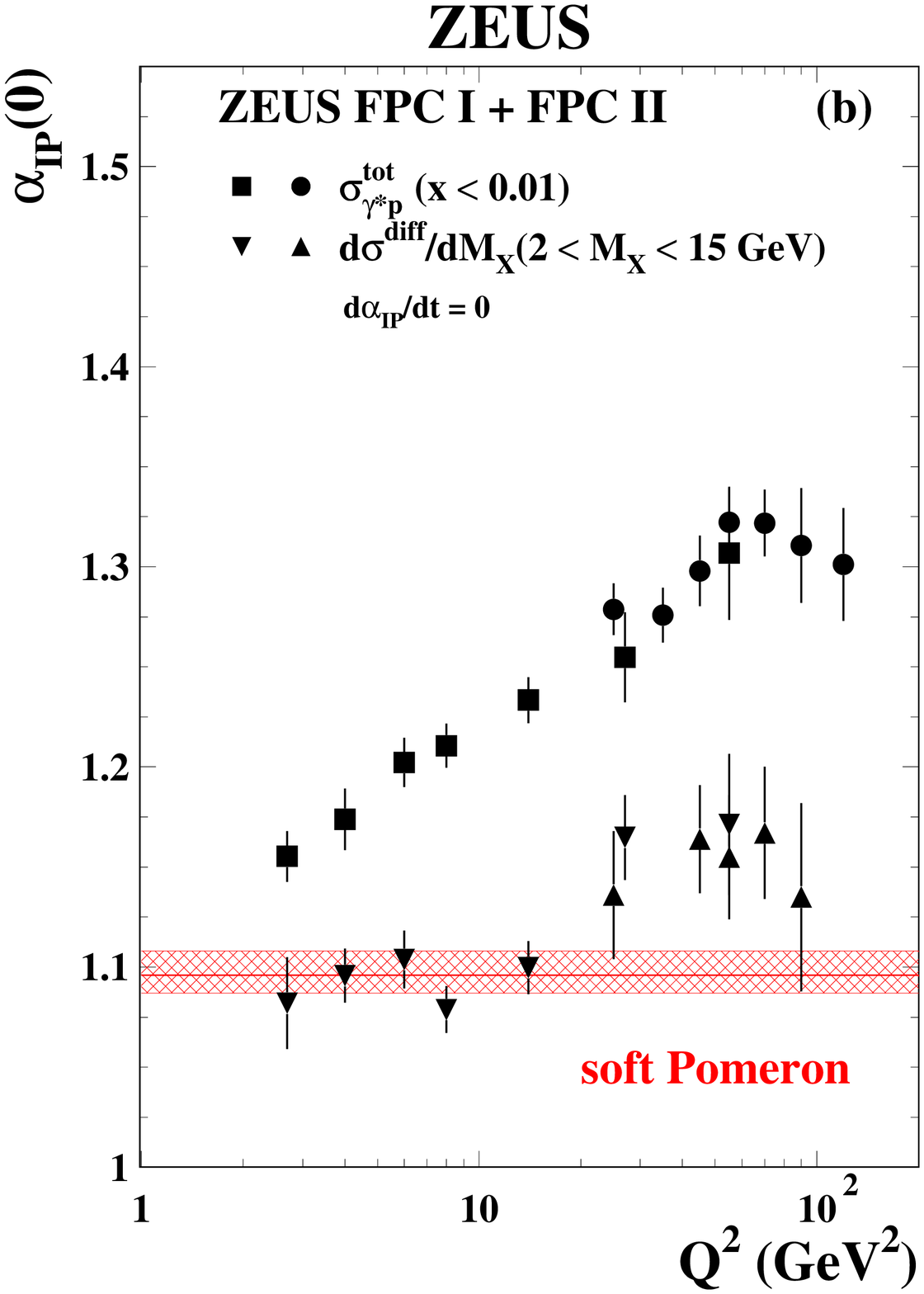,height=85mm}
    \psfig{figure=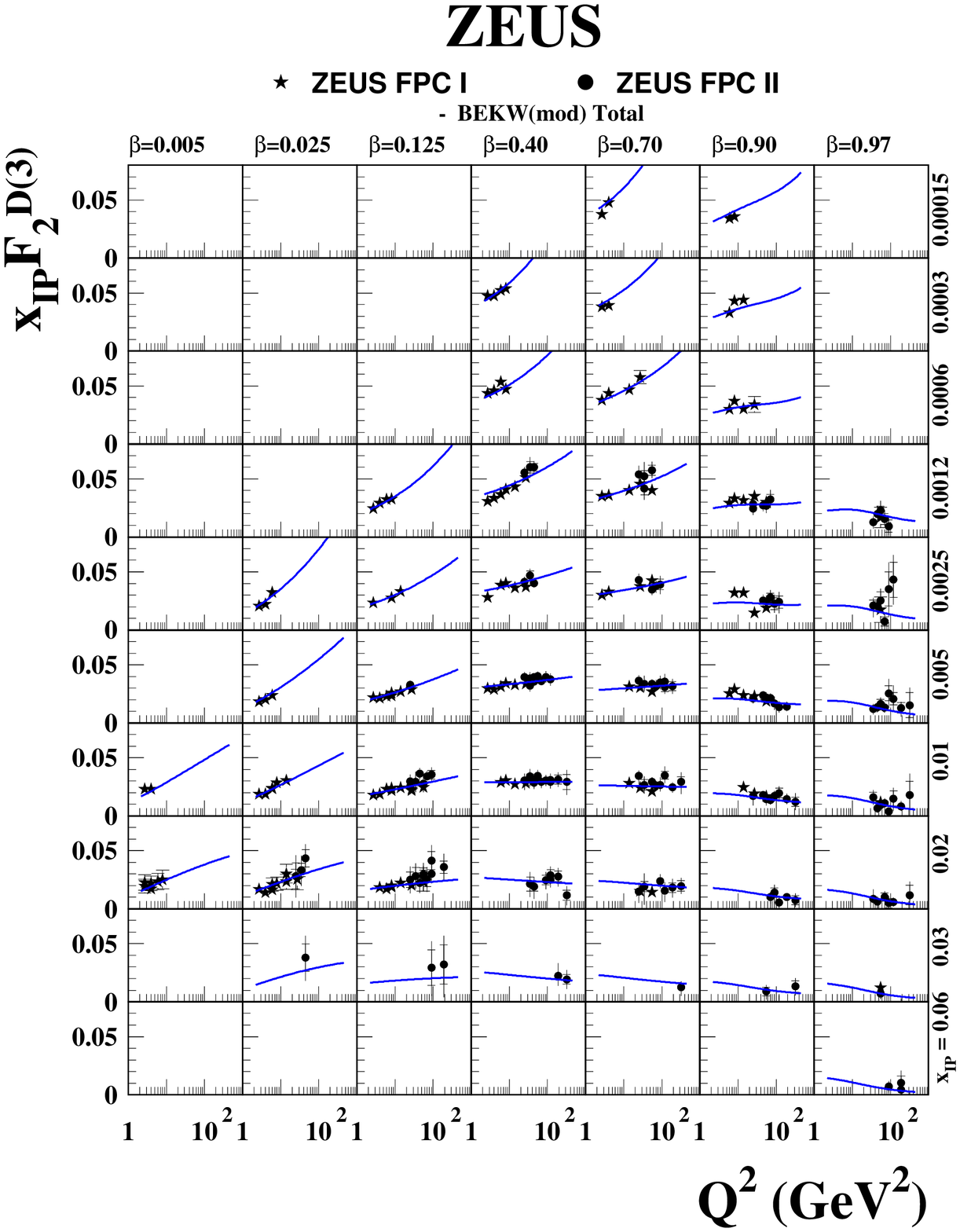,height=100mm}
  \end{center}
  \caption{(left) The intercepts of the Pomeron trajectory, 
    $\alpha ^{\rm tot} _{\pom}(0)$ and $\alpha ^{\rm diff} _{\pom}(0)$ 
    vs. $Q^2$ for $\alpha_{\pom}/dt = \alpha_{\pom} ^{\prime}=0$ 
    GeV$^{-2}$. (right) $\xpom F_2 ^{\rm D(3)}$ vs.
    $Q^2$ for different values of $\beta$ and $\xpom$. The curves are 
    the results of BEKW(mod) fit. Data are shown from the ZEUS FPC I and FPC II
     analyses.}
  \label{fig:FPCII}
\end{figure}

Highlights are presented mainly from the ZEUS diffractive deep inelastic
scattering (DIS) results 
obtained with the $M_X$ method which were recently  published~\cite{FPCII}.
They have been obtained
with data taken in 1998-1999 for $2.2 <Q^2 < 80$ GeV$^2$ (FPC I) 
and in 1999-2000 for $25<Q^2<320$ GeV$^2$ (FPC II). 
The diffractive cross section, 
$d \sigma ^{\rm diff} _{\gamma ^{\ast}p \to XN}/dM_X$ with $M_N < 2.3$ GeV was 
studied as a function of the hadronic centre-of-mass $W$, of the mass $M_X$ of 
the diffractively produced system $X$ and for different $Q^2$ values. The
agreement between both results is observed within the systematic errors. 
For $M_X < 2$ GeV, the diffractive cross section is rather
constant with $W$ while at higher $M_X$, a strong rise with $W$ is observed for all values of $Q^2$. 
The intercept of the Pomeron trajectory was calculated using the $W$ 
dependence of
the diffractive cross section.  
As shown in Fig.~\ref{fig:FPCII}-(left), 
for $Q^2 < 20$ GeV$^2$, $\alpha ^{\rm diff} _{\pom}(0)$ is
compatible with the soft-Pomeron value, while a substantial rise with $Q^2$
above the soft-Pomeron value is observed for $Q^2 > 30$ GeV$^2$.  The
$\alpha ^{\rm diff} _{\pom}(0)$ values are lower than those from the total
$\gamma^{\ast}p$ process, 
with $[\alpha ^{\rm diff} _{\pom}(0) -1]/[\alpha ^{\rm tot} _{\pom} (0) - 1] \approx 0.5 - 0.7$. Therefore, these processes cannot be described 
simultaneously by a single Pomeron.
The ratio of the diffractive cross section to the total $\gamma ^{\ast}p$ 
cross section, $r=\sigma ^{\rm diff} (0.28 < M_X < 35$ GeV, $M_N < 2.3$ 
GeV$)/\sigma ^{\rm tot}$ is independent of $W$ for fixed $Q^2$. At $W=220$ GeV, 
this ratio decreases  $\propto \ln(1+Q^2)$ from 15.8 $\%$ at $Q^2 =4$ GeV$^2$
to 5.0 $\%$ at $Q^2 = 190$ GeV$^2$. 


The diffractive structure function of the proton, $F_2 ^{\rm D(3)}$ is 
parametrized in terms of $Q^2$, the momentum fraction, 
$\xpom = (M_X ^2 + Q^2)/(W^2 + Q^2)$ of the 
proton carried by the Pomeron exchanged in the $t$-channel, 
and the fraction of the Pomeron 
momentum carried by the struck quarks, $\beta = Q^2/(M_X ^2 + Q^2)$. 
If $F_2 ^{\rm D(3)}$ is interpreted in terms of quark densities, 
it specifies the 
probability to find, in a proton undergoing a diffractive reaction, a quark
carrying a fraction $x=\beta \xpom$ of the proton momentum.
The $Q^2$ dependence of $\xpom F_2 ^{D(3)}$ for fixed $\beta$ and 
$\xpom$ is shown in Fig.~\ref{fig:FPCII}-(right). The data are dominated by 
positive scaling violations proportional to $\ln Q^2$ for  
$\xpom \beta = x < 1 \cdot 10 ^{-3}$, while for $x \ge 5 \cdot 10^{-3}$ 
negative violation is observed.
For fixed $\beta$, the $Q^2$ dependence of $\xpom F_2 ^{\rm D(3)}$ changes
with $\xpom$. This is inconsistent with 
the assumption of Regge factorisation, that 
$\xpom F_2 ^{D(3)} (\beta, \xpom, Q^2)$ factorises into a term that depends
only on $\xpom$ and a second term that depends only on $\beta$ and $Q^2$.
The comparison with the H1 LRG measurement~\cite{H1_Etamax} shows fair 
agreement for the $Q^2$ dependence except for a few kinematic regions. 



\section{Diffractive parton densities}
\vspace*{-0.1cm}
\begin{figure}
  \begin{center}
    \psfig{figure=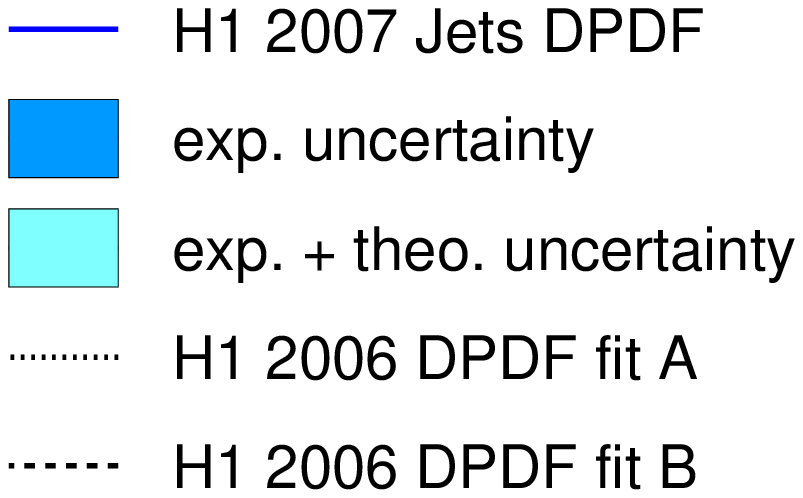,height=19mm}
    \psfig{figure=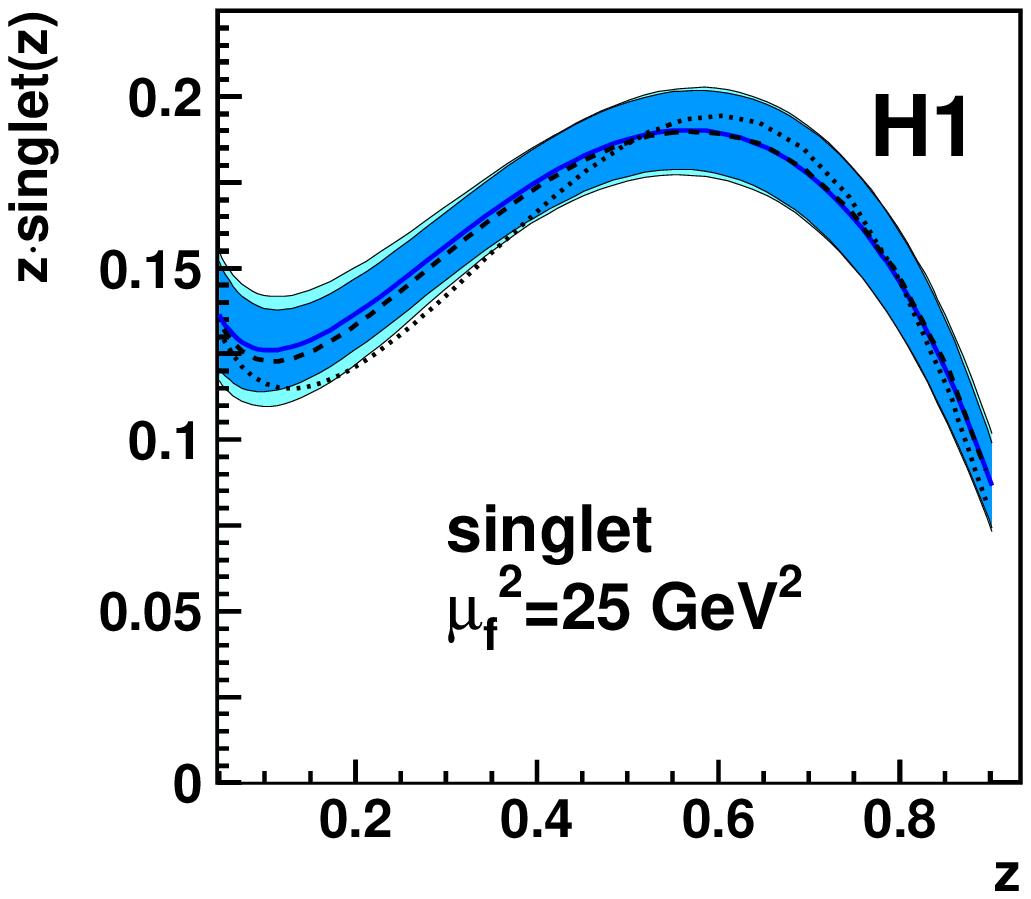,height=30mm}
    \psfig{figure=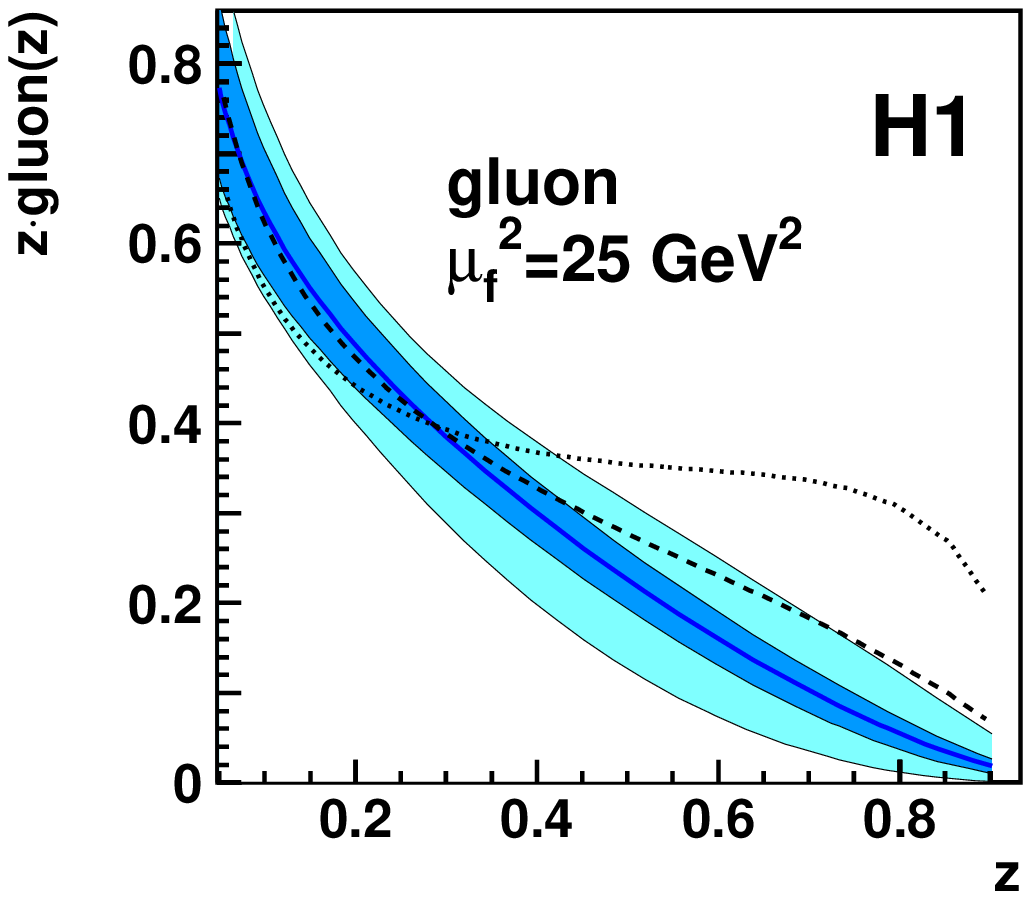,height=30mm}
    \psfig{figure=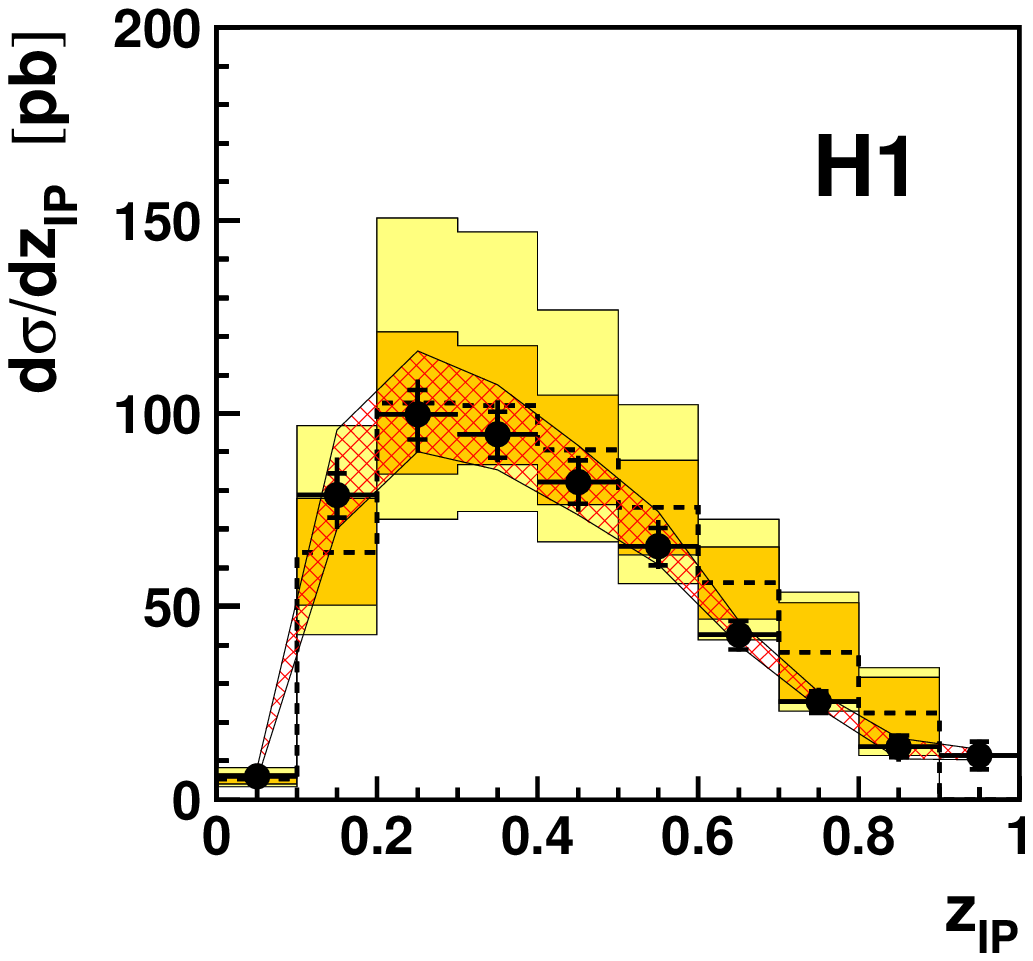,height=33mm}
  \end{center}
  \caption{The diffractive quark density (left) and the diffractive 
    gluon density (middle) vs. $z$ 
    for the squared factorisation scale
    $\mu ^2 _f = 25$ GeV$^2$. (right) Differential cross section for 
    diffractive dijets measured by H1 experiment in DIS vs.
    $\zpom$ compared to NLO prediction using the H1 2006
    DPDF fit B (dashed line). The bands indicate the parton density
    and hadronisation uncertainties (dark-shaded), and the scale
    uncertainty (shaded). } 
  \label{fig:jetpdf}
\end{figure}

%
In Quantum Chromodynamics (QCD), the diffractive DIS processes can be 
described
by a convolution of universal diffractive parton distributions (DPDFs) and
process-dependent coefficients.
The DPDFs have been determined through fits
to the inclusive diffractive data at HERA using the DGLAP evolution equations.
For a parametrisation of the $\xpom$ dependence of the DPDFs, the proton 
vertex factorisation is adopted such that the DPDFs are factorised into 
a term related to the pomeron flux factor depending on $\xpom$ and $t$, and 
a term related to the partonic structure of Pomeron depending on 
$x$(or $\beta$) and $Q^2$.  
In QCD factorisation, next-to-leading order (NLO) QCD calculations based on 
DPDFs could predict the production rates of more
exclusive diffractive processes such as dijet and open charm production. 

The DPDFs extracted from H1 LRG measurement~\cite{H1_Etamax} with $Q^2 > 8.5$ 
GeV$^2$, $M_X \ge 2$ GeV and $\beta \le 0.8$ show that
gluons carry $\approx$ 70 $\%$ of the momentum of the diffractive exchange.
At large $z$ which is the longitudinal
momentum fraction carried by the relevant parton, the diffractive quark 
density remains well constrained, whereas the sensitivity to the gluon 
density becomes increasingly poor and the systematic uncertainties also are
high. It results from the fact that the results from the gluon density at large $z$ are determined
by the inclusive diffractive data at lower $z$ coupled with the chosen 
parametrisation.   
The fit (referred to as `H1 2006 DPDF fit B') was repeated using the 
different parametrisation of the gluon density.
A steeper fall-off in the gluon density at high $z$ is obtained for fit B
than for fit A (see Fig.~\ref{fig:jetpdf}-(middle)), while the quark
densities agree within the uncertainties (see Fig.~\ref{fig:jetpdf}-(left)). 

Dijet production is dominated by the boson-gluon fusion process, 
where a hard collision of a virtual photon and a gluon produces a 
high-$p_{\rm T}$ $q\bar{q}$ pair. Therefore, the diffractive dijet data
are most directly sensitive to the gluonic content of the diffractive 
exchange. 
The prediction based on the H1 2006 DPDF fit B describes the diffractive dijet
data in DIS well (see Fig.~\ref{fig:jetpdf}-(right)), while the H1 2006 DPDF
fit A shows the overestimation comparing with the data in the region of 
high $\zpom$ where $\zpom$ is the fraction of the momentum of the diffractive exchange 
carried by the parton entering the hard interaction. 

A new set of diffractive parton distribution functions (referred to as `H1 2007 
Jets DPDF') has been obtained through
a simultaneous fit to the diffractive inclusive and dijet cross 
sections~\cite{H1_DijetDIS}.
This allows for a precise determination of both the diffractive quark
and gluon distributions in the range $0.05 < \zpom < 0.9$. In particular,
the precision on the gluon density at high momentum fraction is improved
compared to previous extractions. As shown in Fig.~\ref{fig:jetpdf}-(left)
and (right), it is compatible with the H1 2006 DPDF fit B.


\begin{figure}
  \begin{center}
    \psfig{figure=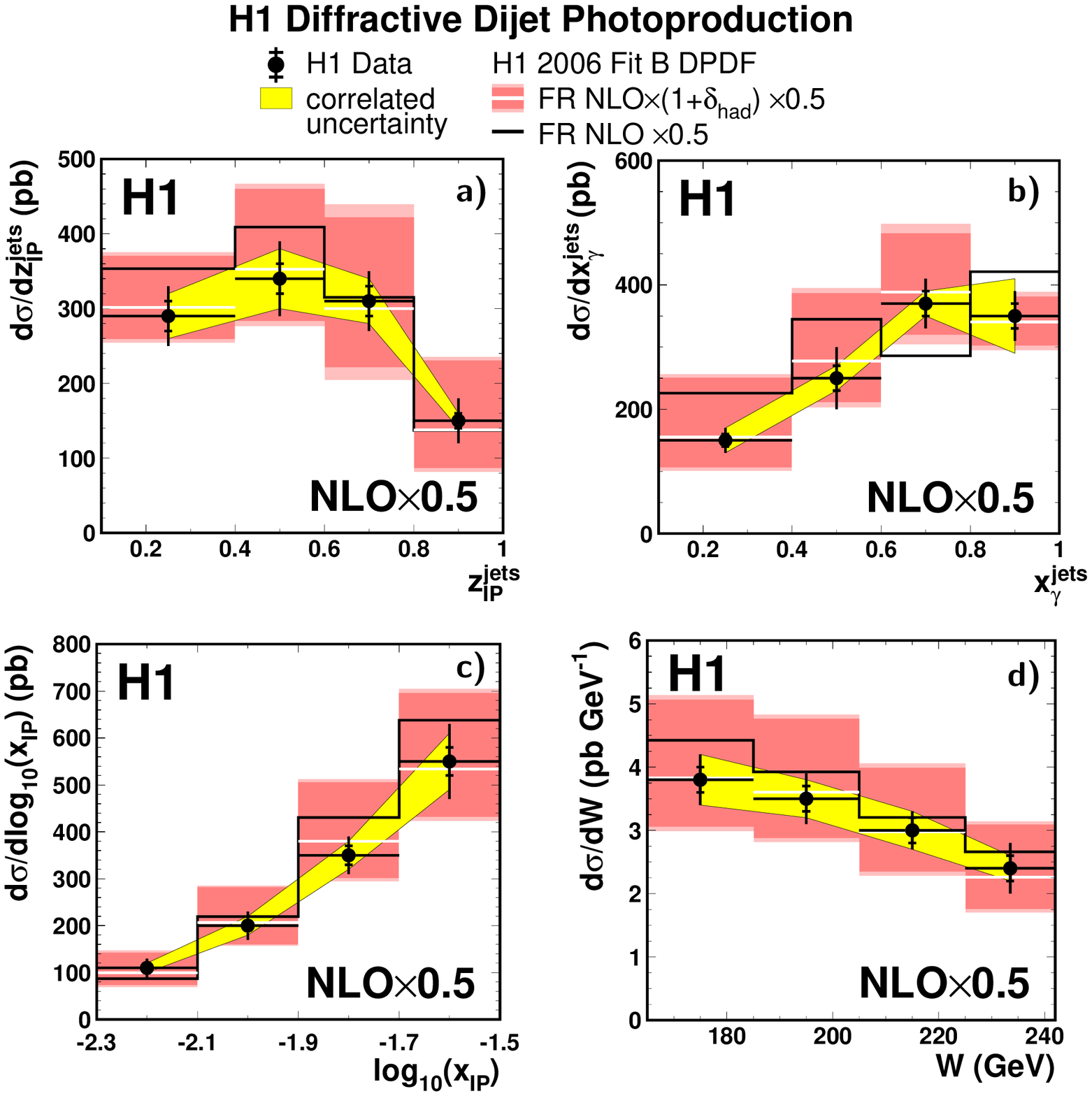,height=70mm}
    \psfig{figure=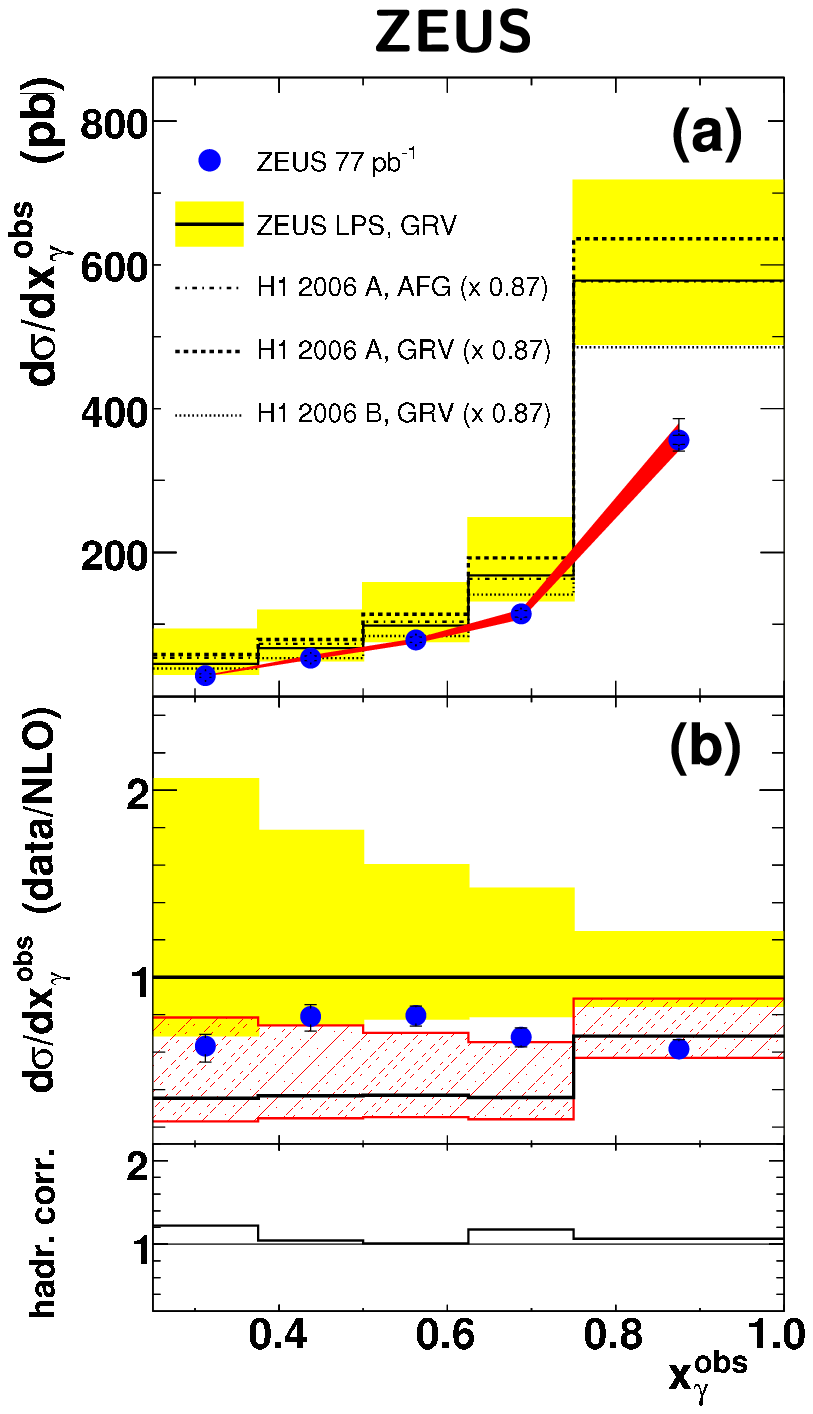,height=70mm}
  \end{center}
  \caption{(left) Differential cross sections for diffractive dijets measured
    by H1 in the 
    photoproduction vs. $\zpomjet$,
    $x^{\rm jets} _{\gamma}$,  log$_{10}(\xpom)$ and $W$. The NLO prediction
    from the H1 2006 DPDF fit B scaled by a normalisation factor 0.5
    is shown. 
    (right)-(a) Differential cross sections for diffractive dijets measured by 
    ZEUS in the photoproduction vs. $x^{\rm obs} _{\gamma}$ with
    NLO QCD predictions. (b) The ratio of the cross sections to the prediction.
    Note that $x_{\gamma}$ is the fraction of the momentum of the photon
    carried by the parton in the hard scattering.}
  \label{fig:jetpdf_PHP}
\end{figure}

The diffractive dijet cross section in DIS agrees well with the QCD 
predictions, supporting the QCD factorisation for 
DIS~\cite{ZEUS_DijetDIS,H1_DijetDIS}.
Processes in which a real photon participates directly in the
hard scattering are expected to be similar to the DIS of highly virtual 
photons.  By contrast,  processes in which the photon is first resolved
into partons which then initiate the hard scattering resemble hadron-hadron
scattering. Gluon-gluon and gluon-quark final states, which are present in the
equivalent $p \bar{p}$ collisions but negligible in DIS, are accessible
via resolved photon processes in hard photoproduction. 
As shown in Fig.~\ref{fig:jetpdf_PHP}-(left), NLO calculations based on the H1 
2006 DPDF fit B overestimate the measured cross section of H1 diffractive
dijets in photoproduction~\cite{H1_DijetPHP}. The ratio
of measured cross section to NLO prediction is a factor  $0.5 \pm 0.1$ 
smaller than the same ratio in DIS, indicating a clear break-down of QCD 
factorisation. However, Fig.~\ref{fig:jetpdf_PHP}-(right) shows that the ZEUS 
diffractive dijet data in photoproduction~\cite{ZEUS_DijetPHP} 
are compatible with QCD factorisation within the large uncertainties of the NLO
calculations. While the ZEUS dijet events required two jets with a transverse
jet energy above $E_T ^{\rm jet1\:(2)} > 7.5$ (6.5) GeV, the H1 events
required two jets with lower $E_T$ like $E_T ^{\rm jet1\:(2)} > 5$ (4) GeV.
Therefore, the $E_T$ distribution of jets may not be well produced by NLO. 
 
%
\vspace*{-0.1cm}
\section*{Acknowledgments}
\vspace*{-0.2cm}
\small
The submitted manuscript has been created by UChicago Argonne, LLC, Operator of Argonne National Laboratory (``Argonne''). Argonne, a U.S. Department of Energy Office of Science laboratory, is operated under Contract No. DE-AC02-06CH11357. The U.S. Government retains for itself, and others acting on its behalf, a paid-up nonexclusive, irrevocable worldwide license in said article to reproduce, prepare derivative works, distribute copies to the public, and perform publicly and display publicly, by or on behalf of the Government.
\normalsize
%


\end{document}